\documentstyle[aps,preprint,psfig,epsf,floats]{revtex}
\begin{document}
\tighten
\draft
%
%
\title{Phase Separation and Coarsening in One-Dimensional Driven
  Diffusive Systems: Local Dynamics Leading to Long-Range Hamiltonians}
\author{M. R. Evans$^1$, Y. Kafri$^2$, H. M. Koduvely$^2$ and 
D. Mukamel$^2$\\
$^1${\it Department of Physics and Astronomy, University of Edinburgh,
  Mayfield Road, Edinburgh EH9 3JZ, U.K.} \\
$^2${\it Department of Physics of Complex Systems, The Weizmann
  Institute of Science, Rehovot 76100, Israel}
}
\date{February 24, 1998}
\maketitle
\begin{abstract}
A driven system of three species of particle diffusing on a
ring is studied in detail. The dynamics is {\it local} and conserves
the three densities. A simple argument suggesting that the model
should phase separate and break the translational symmetry is
given. We show that for the special case where the three densities are
equal the model obeys detailed balance and the steady-state
distribution is governed by a Hamiltonian with asymmetric long-range
interactions. This provides an explicit demonstration of a simple
mechanism for breaking of ergodicity in one dimension. The steady
state of finite-size systems is studied using a generalized matrix
product ansatz. The coarsening process leading to phase separation is
studied numerically and in a mean-field model. The system exhibits
slow dynamics due to trapping in metastable states whose number is 
exponentially large in the system size. The typical domain size
is shown to grow logarithmically in time. Generalizations to a larger
number of species are discussed.

\end{abstract}



\section{Introduction}
Collective phenomena in systems far from thermal equilibrium have been
of considerable interest in recent years \cite{NESM}. Unlike systems
in thermal equilibrium where the Gibbs picture provides a theoretical
framework within which such phenomena can be studied, here no such
framework exists and one has to resort to studies of specific models
in order to gain some understanding of the phenomena involved.

One class of such models is driven diffusive systems (DDS)
\cite{KLS,SZ}. Driven by an external field these systems do not
generically obey detailed balance so that the steady state has
non-vanishing currents. Theoretical studies of DDS have revealed basic
differences between systems in thermal equilibrium and systems far
from thermal equilibrium. For example, it is well known that one
dimensional ($1d$) systems in thermal equilibrium with short-range
interactions do not exhibit phenomena such as phase transitions,
spontaneous symmetry breaking (SSB) and phase separation (except in
the limit of zero temperature or in the context of long-range
interactions) \cite{Landau}. In contrast, some examples of noisy $1d$
DDS with local dynamics have been found to exhibit such phenomena.

One example of a noisy system which exhibits SSB in $1d$ is the
asymmetric exclusion model of two types of charge studied in
\cite{EFGM1,EFGM2}.  In this model, two types of charge are biased to
move in opposite directions on a $1d$ lattice with open ends. The
charges interact via a hard-core interaction, and are injected at one
end of the lattice and ejected at the other end. This model is
symmetric under the combined operations of charge conjugation and
parity (PC symmetry). However, this symmetry is broken in the steady
state, where the currents of the two charges are not equal. The reason
for symmetry breaking in this model lies to some extent in the open
boundaries. Other examples of models in which there is SSB in $1d$
have also been found in the context of cellular automata \cite{Gacs}
and surface growth \cite{JKDW,AEHM}. In the latter, SSB was due to the
fact that one of the rates for a local dynamical move in the models is
zero. Once this zero rate changes to a non-zero rate SSB disappears.

A closely related problem to spontaneous symmetry breaking, is that of
{\it phase separation} in $1d$ noisy systems. This has been observed
in driven diffusive models with inhomogeneities, such as defect sites
\cite{defect} or particles \cite{DJLS}. In these models it has been
found that macroscopic regions of high densities are formed near the
defect, much like a high density of cars behind a slow car in a
traffic jam \cite{Evans,KF}. Here the phase separation is triggered by
the defects. It is of interest to study whether phase separation can
occur in $1d$ noisy homogeneous systems such as on a ring geometry
with no defects, where {\it all} possible local transition rates which
are consistent with the symmetry and conservation laws of the model
are non vanishing. Recently, Lahiri and Ramaswamy have introduced a
lattice model in the context of sedimenting colloidal crystals, where
phase separation is found to take place without any inhomogeneities
\cite{RLSR}. In this model, there are two rings coupled to each other
and particles on each ring undergo an asymmetric exclusion
process. The hopping rate between sites $i$ and $i+1$ on each ring
depends on the occupation at the $i$th site on the other ring. However,
this model is studied mainly using Monte Carlo simulations and no
analytical results are available so far.

In a recent Letter \cite{EKKM} we introduced a simple three-species
driven diffusive model exhibiting phase separation and spontaneous breaking of
the translational symmetry on a ring. In the model nearest-neighbor
particles exchange with given rates and the numbers of each species are
conserved under the dynamics.  The rates of all  local dynamical moves that
obey the conservation laws are non zero.  An argument indicating that
generically the system phase separates, thus breaking the translational 
invariance, was given for the case when none of the species
of particle has zero density. In the special case of equal
number of particles of each type, it was shown that the local dynamics
obeys detailed balance with respect to a long-range asymmetric
(chiral) Hamiltonian.  In this special case, using the Hamiltonian, we
have found the steady state of the model exactly and have been able to
prove the existence of phase separation analytically.

The existence of a Hamiltonian for this special case is of interest in
the light of speculation that non-equilibrium systems exhibiting
generic long-range correlations might be described by effective
Hamiltonians containing long-range interactions \cite{BR,AE}. Here we
explicitly demonstrate that for the special case where the three
densities are equal the model is exactly described by a long-range
asymmetric Hamiltonian. The model not only has long-range correlations
but has generic long-range {\it order}. The mechanism found in this 
study suggests that systems with dynamical rules defined completely 
locally and {\it a priori}
without respect to any Hamiltonian, may have a steady state where the
configuration space is sampled according to a measure that is intrinsically
global. The Hamiltonian also allows us to identify the analog of a
temperature in the microscopic dynamics as related to the drive of the
system; for zero drive, that is symmetric diffusion of the particles,
the effective temperature is infinite and phase separation is lost.

We note that a related but distinct three-species model has recently
been introduced by Arndt {\it et al}. This model also exhibits phase
separation\cite{Rit}.

In the present work we analyze in detail the $M=3$ species model which
was introduced in \cite{EKKM} and then generalize it to larger $M$.
We provide the complete proof of phase separation which follows from the
exact calculation of the partition sum in the thermodynamic limit.  We
also provide numerical evidence of phase separation in the general
case where the densities of the three particles are not equal.

In order to study the coarsening process Monte-Carlo simulations are
 performed. However, simulation of the microscopic model is hampered by 
slow dynamics which makes it difficult to
access the scaling regime. The system becomes trapped in metastable 
states comprising
several domains of each type of particle. The number of metastable
 states is
exponentially large in the system size. The lifetimes of the
metastable states increase exponentially with the average domain size
as the fully phase separated state is approached. Thus the model
provides an example of slow dynamics in a system without any quenched
disorder \cite{Ritort}.

To ameliorate the difficulty of numerically
studying such slow dynamics we employ a toy model wherein it is the
domains that are updated rather than the individual particles.  This
allows the long-time scaling behavior of the domain size to be
investigated and to confirm a logarithmic growth of the average domain
size with time. The toy model also affords a mean-field solution for
the long-time dynamical behavior, that again confirms the scaling
behavior.

Returning to the case of equal numbers of particles of different
species it is of interest to investigate the steady-state behavior in
finite-size systems.  We have found it convenient to do this by
employing a matrix product technique previously used to solve the
steady state of asymmetric exclusion processes \cite{DEHP}. However, 
in the case of three species the simplest form of this technique
 \cite{DJLS,DEHP,Evans} is applicable only 
to a limited class of systems
\cite{AHR}.  For the present model we generalize the
matrix product to a product of rank 6 tensors and write the steady
state by taking an appropriate contraction. The partition sum and
steady-state correlation functions can be conveniently computed
numerically using this tensor product ansatz.

The paper is organized as follows: in section II we define the model
introduced in \cite{EKKM} and we present an argument which indicates
that the system should phase separate as long as none of the species
of particles has zero density. In section III we study the special
case where the model satisfies detailed balance and explicitly write
down the steady-state weight for the three-species model. The
existence of phase separation in the model for any non-infinite
temperature is proved analytically by calculating some bounds on the
two-point correlation functions.  Section IV contains numerical
evidence for phase separation in the general case where the densities
of the three species of particles are not equal. The toy model, which
facilitates efficient Monte Carlo simulations, is used to study the
dynamics of phase separation. A mean-field analysis of this toy
model is presented, the details being left to Appendix A.  In Section
V we present results for finite systems obtained via the tensor
product ansatz. Using these results we study finite-size scaling in
the system.  In section VI we address phase separation in systems with
more than three species of particle and a proof of detailed balance
for special cases is given in Appendix B. We conclude in section VII
and discuss some open questions.

\section{Definition of the model}
We start by defining a three-species model which exhibits phase separation
in $1d$. Consider a one-dimensional, ring-like (periodic) lattice of length $N$
where each site is occupied by one of the three types of particles,
$A$, $B$, or $C$. The model evolves under a random sequential update
procedure which is defined as follows: at each time step two
neighboring sites are chosen randomly and the particles at these sites
are exchanged according to the following rates
\begin{equation}
\label{eq:dynamics}
\begin{picture}(130,37)(0,2)
\unitlength=1.0pt
\put(36,6){$BC$}
\put(56,4) {$\longleftarrow$}
\put(62,0) {\footnotesize $1$}
\put(56,8) {$\longrightarrow$}
\put(62,13) {\footnotesize $q$}
\put(80,6){$CB$}
\put(36,28){$AB$}
\put(56,26) {$\longleftarrow$}
\put(62,22) {\footnotesize $1$}
\put(56,30) {$\longrightarrow$}
\put(62,35) {\footnotesize $q$}
\put(80,28){$BA$}
\put(36,-16){$CA$}
\put(56,-18) {$\longleftarrow$}
\put(62,-22) {\footnotesize $1$}
\put(56,-14) {$\longrightarrow$}
\put(62,-9) {\footnotesize $q$}
\put(80,-16){$AC$.}
\end{picture}
\end{equation}
\vspace{0.4cm}

\noindent The particles thus diffuse asymmetrically around the ring. The
dynamics conserves the number of particles, $N_A , N_B$ and  $N_C$ of
the three species.

The $q=1$ case is special. Here the diffusion is symmetric and every
local exchange of particles takes place with the same rate as the
reverse move. The system thus obeys detailed balance reaching a steady
state in which all microscopic configurations (compatible with the
number of particles $N_A , N_B$ and $N_C$) are equally probable. This
state is homogeneous, and no phase separation takes place. We now
present a simple argument suggesting that for $q\neq 1$ the steady
state of the system is not homogeneous in the thermodynamic limit. For
simplicity the case $q < 1$ is examined. As a result of the bias in
the exchange rates an $A$ particle prefers to move to the left inside
a $B$ domain and to the right inside a $C$ domain. Similarly the
motion of $B$ and $C$ particles in foreign domains is biased. Consider
the dynamics starting from a random initial configuration. The
configuration is composed of a random sequence of domains of $A$, $B$,
and $C$ particles. Due to the bias a local configuration in which an
$A$ domain is placed to the right of a $B$ domain is unstable and the
two domains exchange places on a relatively short time scale which is
linear in the domain size.  Similarly, $AC$ and $CB$ domains are
unstable too. On the other hand $AB$, $BC$ and $CA$ configurations are
stable and long lived. Thus after a relatively short time the system
reaches a state of the type $\ldots AAABBCCAABBBCCC\ldots$ in which
$A,B$ and $C$ domains are located to the right of $C,A$ and $B$
domains, respectively. The evolution of this state takes place via a
slow diffusion process in which, for example, the time scale for an
$A$ particle to cross an adjacent $B$ domain is $q^{-l}$, where $l$ is
the size of the $B$ domain.  The system therefore coarsens and the
average domain size increases with time as $\ln t/\vert \ln q\vert$
\cite{SHS}.  Eventually the system phase separates into three domains
of the three species of the form $A\ldots AB\ldots BC\ldots C$.

In a finite system the phase separated state may further evolve and
become disordered due to fluctuations.  However, the time scale for
this to happen grows exponentially with the system size. For example
it would take a time of order of $q^{-min\{N_B,N_C\}}$ for the $A$
domain in the totally phase separated state to break up into smaller
domains. Hence in the thermodynamic limit, this time scale diverges
and the phase separated state remains stable provided the density of
each species is non-zero. Note that there are always small
fluctuations about a totally phase separated state. However, these
fluctuations affect the densities only near the domain
boundaries. They result in a finite width for the domain walls. The
fact that any phase separated state is stable for a time exponentially
long in the system size amounts to a breaking of the translational
symmetry.

Since the exchange rates are asymmetric, the system generically
supports a particle current in the steady state. To see this, consider
the $A$ domain in the phase separated state.  An $A$ particle near the
$\ldots AB \ldots$ boundary can traverse the entire $B$ domain to the
right with an effective rate proportional to $q^{N_B}$.  Once it
crosses the $B$ domain it will move through the $C$ domain with rate
$1-q$. Similarly an $A$ particle near the $\ldots CA \ldots$ boundary
can traverse the entire $C$ domain to the left with a rate
proportional to $q^{N_C}$. Once the domain is crossed it moves through
the $B$ domain with rate $1-q$. Hence the net $A$ particle current is
of the order of $q^{N_B} - q^{N_C}$. Since this current is
exponentially small in system size, it vanishes in the thermodynamic
limit.  For the case of $N_A=N_B=N_C$, this argument suggests that the
current is strictly zero for any $N$. In sections III and V we study
this case in detail.

The arguments presented above suggesting phase separation for $q<1$
may be easily extended to $q>1$. In this case, however, the phase
separated state is $BAC$ rather than $ABC$. This may be seen by noting
that the dynamical rules are invariant under the transformation $q
\rightarrow 1/q$ together with $A \leftrightarrow B$.

\section{Special Case $N_A=N_B=N_C$}
In this section we show that the dynamics (\ref{eq:dynamics}), for the
special case $N_A=N_B=N_C$, satisfies detailed balance. The
corresponding Hamiltonian, which determines the steady-state
distribution, is found to have long-range asymmetric
interactions. Using this Hamiltonian, we calculate analytically the
partition sum and bounds on the correlation functions in the
thermodynamic limit. These are then used to prove the existence of
phase separation in the model. Later, in section \ref{matrixans} we
study finite systems for this case and the approach to the
thermodynamic limit.

\subsection{Detailed Balance}
The general argument presented in the previous section suggests that
for the special case $N_A=N_B=N_C$, the steady state carries no
current for any system size. We demonstrate this explicitly by showing
that the {\it local dynamics} of the model satisfies detailed balance
with respect to a {\it long-range asymmetric} Hamiltonian ${\cal H}$.

We define the occupation variables $A_i,B_i$ and $C_i$ as follows:
\begin{eqnarray}
\label{ocup}
A_i = \left\{ \begin{array}{ll}
            1 & \text{if site $i$ is occupied by an $A$ particle} \\
            0 & \text{otherwise. }
            \end{array}
       \right.
\end{eqnarray}
The variables $B_i$ and $C_i$ are defined similarly. Clearly the
relation $A_i+B_i+C_i = 1$ is satisfied. A microscopic configuration
is thus described by a set $\{X_i\}=\{A_i,B_i,C_i\}$.  Using these
variables, we will show that the Hamiltonian ${\cal H}$ and the
steady-state distribution $W_N$ corresponding to the dynamics
(\ref{eq:dynamics}) for the case $N_A=N_B=N_C=N/3$ are given by
\begin{equation}
\label{eq:Hamilton}
{\cal H}(\lbrace X_i \rbrace) =\sum_{i=1}^{N-1}
\sum_{j=i+1}^{N}[C_iB_j-C_iA_j+B_iA_j] \; ,
\end{equation}
\begin{equation}
\label{eq:weight}
W_N(\{X_i\}) = Z_N^{-1}q^{{\cal H}(\lbrace X_i \rbrace)} \;.
\end{equation}
Here $Z_N$ is the partition sum given by $\sum q^{{\cal H} (\lbrace
X_i \rbrace)}$, where the sum is over all configurations in which
$N_A=N_B=N_C$. Note that the Hamiltonian ${\cal H}$ does not
determine the dynamics of the system, it just governs the steady-state
distribution as given in (\ref{eq:Hamilton}) and
(\ref{eq:weight}). Eq. (\ref{eq:weight}) suggests that $q$
serves as a temperature variable with $kT=-1/\ln q$. Thus, $q \to 1$ 
is the infinite-temperature limit.
 The Hamiltonian (\ref{eq:Hamilton}) is written in a
form which is not manifestly translationally invariant. However,
careful examination reveals that when the relation $N_A=N_B=N_C$ is
taken into account, the Hamiltonian as given by (\ref{eq:Hamilton}) is
indeed translationally invariant (see Appendix B). Therefore site $1$
may be chosen arbitrarily. An expression for ${\cal H}$ which is
manifestly translationally invariant will be derived at the end of
this section.

Note that
\begin{equation}
\sum_{i=1}^{N-1}\sum_{j=i+1}^{N} (C_iA_j+A_iC_j) = (N/3)^2 
\end{equation}
since the LHS yields the number of $CA$ (and $AC$) pairs in the
system.  Using this relation the Hamiltonian may also be written in a
form where the cyclic symmetry is more apparent:
\begin{equation}
{\cal H}(\lbrace X_i \rbrace)  = 
\sum_{i=1}^{N-1}\sum_{j=i+1}^{N}[C_iB_j+A_iC_j+B_iA_j]-(N/3)^2 \; .
\end{equation}

The proof of Eqs. (\ref{eq:Hamilton},\ref{eq:weight}) is
straightforward. This is done by considering a nearest-neighbor
particle exchange and verifying that detailed balance is satisfied
with respect to (\ref{eq:weight}). We start by considering nearest-neighbor
 sites in the interior of the lattice, namely pairs other than
$(1,N)$.  For example consider the exchange $AB \rightarrow BA$
taking place at two adjacent sites $k$ and $k+1$, where $k \ne
N$. This exchange results in the contribution of one more $B_iA_j$
term in $\cal H$ and hence the energy of the resulting configuration
is higher by $1$. It is easy to see using Eq. (\ref{eq:weight})
that $q W_N(\ldots AB \ldots) = W_N(\ldots BA \ldots)$, as required by
detailed balance. Similar relations are easily derived for exchange of
$BC$ and $CA$ pairs. Now consider an exchange taking place between sites
$N$ and $1$, say $CA\rightarrow AC$.  According to (\ref{eq:Hamilton})
this exchange  costs an energy of $2N_B-N_A-N_C+1$. Therefore the
exchange satisfies the detailed balance condition $q W_N(A\ldots C) =
W_N(C\ldots A)$ only when $2N_B=N_A+N_C$. Similarly, by considering
the exchanges $AB\rightarrow BA$ and $BC\rightarrow CB$, one deduces
that the detailed balance condition is satisfied for any exchange at
sites $N$ and $1$ as long as $N_A=N_B=N_C$. In Appendix B we consider
the most general nearest-neighbor exchange rates for $M$ species and
arbitrary densities and derive conditions (\ref{condition}) for
exchange rates which satisfy detailed balance.

To write ${\cal H}$ in a manifestly translationally invariant form we
define ${\cal H}_{i_0}(\lbrace X_i \rbrace)$ as the Hamiltonian in
which site $i_0$ is the origin. Namely,
\begin{equation}
\label{hi0}
{\cal H}_{i_0}(\lbrace X_i \rbrace) =   
\sum_{i=i_0}^{N+i_0-2}
\sum_{j=i+1}^{N+i_0-1}[C_iB_j-C_iA_j+B_iA_j] \; ,
\end{equation}
where the summation over $i$ and $j$ is modulo $N$. Summing
(\ref{hi0}) over all $i_0$ and dividing by $N$, one obtains,
\begin{eqnarray}
{\cal H}(\{ X_i \})  & = & \sum_{i=1}^N \sum_{k=1}^{N-1}
(1 - \frac{k}{N}) (C_i B_{i+k} - C_i A_{i+k} + B_i A_{i+k}) 
\label{eq:H_tinvar} \\
& = & \sum_{i=1}^N \sum_{k=1}^{N-1}
(1 - \frac{k}{N}) (C_i B_{i+k} + A_i C_{i+k} + B_i A_{i+k}) - (N/3)^2 \; ,
\end{eqnarray}
where in the summation the value of the site index $(i+k)$ is modulo
$N$. In the Hamiltonian (\ref{eq:H_tinvar}) the interaction is linear
in the distance between the particles, and thus is long-ranged.  The
distance is measured in a preferred direction from site $i$ to site
$i+k$. Moreover it is asymmetric in the sense that $\cal H$ is not
invariant under the parity operation. It is also related to chiral
Hamiltonians \cite{CHIRAL}.

\subsection{Ground States and Metastable States}
Before proceeding further to evaluate the partition sum and some
correlation functions associated with the Hamiltonian
(\ref{eq:Hamilton}) let us make a few observations. The ground state
of the Hamiltonian is given by the fully separated state $A\ldots
AB\ldots BC\ldots C$ and its translationally related states. The
degeneracy of the ground state is thus $N$ and its energy is zero. A
simple way of evaluating the energy of an arbitrary configuration is
obtained by noting that nearest-neighbor (nn) exchanges $AB \rightarrow BA,
BC \rightarrow CB$ and $CA \rightarrow AC$ cost one unit of energy
each while the reverse exchanges result in an energy gain of one
unit. The energy of an arbitrary configuration may thus be evaluated
by starting with the ground state and performing nn
exchanges until the configuration is reached, keeping track of the
energy changes at each step of the way.  The highest energy is $N^2/9$
and it corresponds to the totally phase separated configuration
$A\ldots AC\ldots CB\ldots B$ and its $N$ translations.

In considering the excited states of the Hamiltonian
(\ref{eq:Hamilton}) we note that the model exhibits a set of
metastable states which correspond to local minima of the energy: any
exchange of nn particles results in an increase of
the energy.  In these states no $BA,CB$ and $AC$ nn pairs exist; only
$AB,BC$ and $CA$ nn pairs may be found in addition to $AA$, $BB$, and
$CC$. Any metastable state is thus composed of a sequence of domains
in which $A,B$ and $C$ domains follow $C,A$ and $B$ domains,
respectively. Therefore each metastable state has an equal number of
domains, $s$, of each type with $s=1,...,N/3$.  The $s=1$ case
corresponds to the ground state while $s=N/3$ corresponds to the
$ABCABC\ldots ABC$ state, composed of a total of $N$ domains each of
length $1$. (The total number of domains in an $s$-state is $3s$.)

For calculating the free energy and some correlation functions
corresponding to the Hamiltonian (\ref{eq:Hamilton}) we find it useful
to first derive some bounds for the number ${\cal N} (s)$ of
$s$-states and their energies. In the following such bounds are
presented. They are then used, in the next section, to evaluate the
free energy and correlation functions of the model.

To obtain a bound for ${\cal N} (s)$ we note that the number of ways
of dividing $N/3$ $A$ particles into $s$ domains is ${{N/3 -1} \choose
{s-1}}$.  The number of ways of combining $s$ divisions of each of the
three types of particles is clearly $\left[{N/3-1}\choose
{s-1}\right]^3$. There are at most $N$ ways of placing this string of
domains on a lattice to obtain a metastable state (the number of ways
need not be equal to $N$ since the string may possess some
translational symmetry). One therefore has
\begin{equation}
\label{nmeta}
\left[{N/3-1}\choose {s-1}\right]^3 \le{\cal N}(s) \le N
\left[{N/3-1}\choose {s-1}\right]^3\; .
\label{nMSS}
\end{equation}
Thus, the total number of metastable states is exponential in $N$.

We now consider the energy of the metastable states. It is easy to
convince oneself that among all $s$-states, none has energy lower than
the following configuration,
\begin{equation}
A\ldots AB\ldots BC\ldots CABCABC\ldots ABC \; ,
\end{equation}
where the $3(s-1)$ rightmost domains are of size $1$ and the three
leftmost domains are of size $(N/3-s+1)$ each. The energy of this
state, $E_s$ satisfies the following recursion relation
\begin{equation}
\label{RR}
E_s = E_{s-1} + N/3 - s
\end{equation}
with $E_1 =0$. To see this one notes that the $s$-state may be created
from the $(s-1)$-state by first moving a $B$ particle from the
leftmost $B$ domain across $(N/3-s)$ $C$ particles to the right and
then moving an $A$ particle from the leftmost $A$ domain to the right
across the adjacent $B$ and $C$ domains. The energy cost of these
moves is $(N/3-s)$, yielding (\ref{RR}). The recursion relation
(\ref{RR}), together with $E_1=0$, is then readily solved to give
\begin{equation}
E_s = (s-1) \frac{N}{3} - \frac{s(s-1)}{2} \; .
\label{eMSS}
\end{equation}

The energy of all metastable $s$-states is larger or equal to $E_s$ as
given by Eq.(\ref{eMSS}). In the following Section we use the bounds
(\ref{nMSS}) and (\ref{eMSS}) to calculate the partition sum and some
correlation functions corresponding to the Hamiltonian
(\ref{eq:Hamilton}).
\subsection{Partition Sum}
In this section we prove that, in the large $N$ limit and for
all $q<1$, the partition sum is given by
\begin{equation}
Z_N = N/[(q)_{\infty}]^3 \; ,
\label{psum}
\end{equation} 
where
\begin{equation}
(q)_{\infty}=\lim_{n\rightarrow \infty} (1-q)(1-q^2)\ldots (1-q^n) \; .
\end{equation}
The partition sum for $q>1$ may be obtained by replacing $q$ by $1/q$
in (\ref{psum}). Note that the partition sum is linear and not
exponential in $N$, meaning that the free energy is not
extensive. This is a result of the long-range interaction in the
Hamiltonian and the fact that the energy excitations are localized
near the domain boundaries, as will be shown in the following.

For $q$ close to $1$, $(q)_{\infty}$ has an essential singularity,
\begin{equation}
(q)_{\infty}=e^{-\frac{1}{\ln q} \left[ \pi^2/6 + {\cal O}(1-q) \right]}\;.
\end{equation}
This suggests that extensivity of the free energy could be restored in
the double limit $q \rightarrow 1$ and $N \rightarrow \infty$ with $N
\ln q$ finite. This scaling behavior is only suggestive since
expression (\ref{psum}) may not be valid in this limit. For $q=1$ all
configurations with $N_A=N_B=N_C$ are equally probable, so that the
partition sum is given by $Z_N={{N}\choose {N/3}}{{2N/3} \choose
{N/3}}$ which goes like $3^N$ for large $N$.

It is instructive to first present the proof of (\ref{psum}) for small
values of $q$. This proof will then serve as the basis for the proof
for any $q<1$.

We start by noting that in calculating the partition sum (\ref{psum}),
configurations with energy larger then $aN$ ($a>0$) may be neglected
in the thermodynamic limit. For simplicity we first demonstrate this
for $q<(1/3)^{1/a}$, although later we show it for any $q<1$. The
contribution to the partition sum from these energy states,
$Z_{m>aN}$, is given by
\begin{equation}
\label{hexpsum}
Z_{m>aN}=\sum_{m=aN+1}^{N^2/9}D(m)q^m \; ,
\end{equation}
where $D(m)$ is the number of configurations of energy $m$. Clearly,
the number of possible configurations in the system is bounded crudely
from above by $3^N$. This bound implies
\begin{equation}
\label{bhexpsum}
Z_{m>aN}<\sum_{m=aN+1}^{N^2/9}3^Nq^m \; . 
\end{equation}
Thus, for $q<(1/3)^{1/a}$, the contribution to the partition sum
arising from energies larger than $aN$ is exponentially small in $N$
 and may be neglected in the thermodynamic limit.

\begin{figure}
\center{\psfig{figure=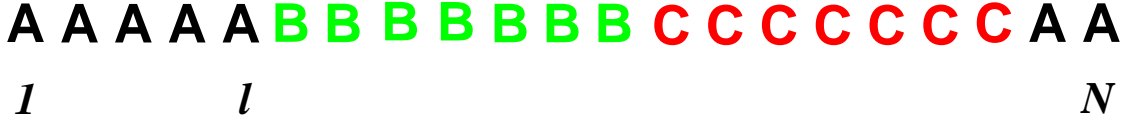,width=16cm}}
\caption{The $l=5$ ground state for an  $N=21$ system.}
\label{ground}
\end{figure}

The calculation of $Z_N$ is thus reduced to calculating a truncated
partition sum in which only energies up to $aN$ are summed over. To
proceed we consider $a \leq 1/3$ and take into account configurations
with energy less than $N/3-1$. This simplifies the calculations
considerably since all configurations with energy $m<N/3-1$ may be
decomposed into $N$ disjoint sets of states, each corresponding to a
unique ground state (see previous section for a discussion of ground
states). We label the sets by $l=1 \ldots N$, the position of the
rightmost $A$ particle in the $A$ domain of the ground state belonging
to the set (see Fig. \ref{ground} for an example of an $l=5$ ground
state). Each state in a specific set can be obtained from the
corresponding ground state by exchanging nearest neighbors so that the
energy always increases along the intermediate states. Note that this
is correct only if excitations of energy less than $N/3-1$ are
considered. This is because not all higher energy states can be
reached by uphill steps from a ground state.

Thus, using translational invariance, the partition
sum can be written as
\begin{equation}
\label{exp}
Z_N=N{\cal Z}_N+e^{-{\cal O}(N)} \; ,
\end{equation}
where ${\cal Z}_N$ is the truncated partition sum of one of the
$N$ sets of configurations.

We proceed to calculate ${\cal Z}_N$. This is done by considering all
the possible energy excitations with energy less than $N/3-1$ above
one ground state. Consider a specific domain boundary, say
$AB$. Excitations of energy $m$ at this boundary can be created by
moving one or more $A$ particles into the $B$ domain (this is
equivalent to moving $B$ particles into the $A$ domain). An $A$
particle moving into the $B$ domain is considered as a walker. The
excitation energy increases linearly with the distance the walker has
moved. Thus, in this picture an excitation of energy $m$ is created by
$1 \leq j \leq m$ walkers, traveling a total distance $m$. The number
of excitations of energy $m$ is then given by the number of ways,
$P(m)$, of partitioning an integer $m$ into the sum of a sequence of
non-increasing positive integers. Taking into account excitations at
all three boundaries, an excitation of energy $m$ in the system is
created by three independent excitations of energy $m_1,m_2$ and $m_3$
at the different domain boundaries such that $m_1+m_2+m_3=m$. The
number of excitations of this form is just given by
$P(m_1)P(m_2)P(m_3)$. Thus, ${\cal Z}_N$ is given by
\begin{equation}
\label{partN}
{\cal
Z}_N=\sum_{m=0}^{N/3-2}q^m\sum_{m_i=0}^mP(m_1)P(m_2)P(m_3)
\delta_{m_1+m_2+m_3,m} \; .
\end{equation}
Taking the thermodynamic limit we obtain
\begin{equation}
\label{thelim}
\lim_{N \rightarrow \infty} {\cal Z}_N=(\sum_{m=0}^{\infty}q^mP(m))^3\;.
\end{equation}
Using a well known result from number theory, attributed to Euler, for
the generating function of $P(m)$ \cite{andrew},
\begin{equation}
\sum_{m=0}^{\infty} q^m P(m) = \frac{1}{(q)_{\infty}}
\end{equation}
 and using (\ref{thelim})
and (\ref{exp}), Eq.(\ref{psum}) is obtained.

So far we have proved that for $q \leq (1/3)^{3}$, Eq.(\ref{psum}) is
exact in the thermodynamic limit. We now extend these results for any
$q<1$. First we have to show that the states ignored in the previous
calculation for $q \leq (1/3)^{3}$ may be ignored for all $q<1$. To do
this we calculate upper and lower bounds on $Z_N$ and show they
converge for large enough $N$.

 For this we have to consider the entire energy spectrum of the
Hamiltonian. Any configuration of the system which is neither a ground
state nor a metastable state can be obtained from at least one ground
state ($s=1$) or a metastable state ($s>1$) as follows: starting from
this $s$-state exchange nearest neighbors such that the energy always
increases along the path until the configuration is reached. In what
follows it is demonstrated that none of the configurations which can
be obtained from  $s$-states, with $s>1$, by the above procedure of
particle exchange, contributes to the partition sum in the
thermodynamic limit.

An upper bound on the partition sum may be calculated as follows:
using the same steps of derivation used for computing ${\cal Z}_N$, it
is straightforward to show that the contribution to the partition sum
from an $s$-state and associated configurations is {\it at most}
$q^{(s-1)N/3 - s(s-1)/2} [(q)_{\infty}]^{-3s}$. The prefactor
$q^{(s-1)N/3 - s(s-1)/2}$ arises from the minimum energy (\ref{eMSS})
of this metastable state. Therefore by considering the contributions
from all the $s$-states and using (\ref{nmeta}) the following bound is
found
\begin{equation}
{\cal Z}_N < N/[(q)_{\infty}]^3 + \sum_{s=2}^{N/3}N {N/3-1\choose s-1}
q^{(s-1)N/3 -s(s-1)/2} [(q_{\infty})]^{-3s}\;.
\label{zupperb1}
\end{equation}
The second term on the RHS represents the contribution from
excitations around the metastable states. Replacing $q^{(s-1)N/3
-s(s-1)/2}$ by an upper bound $q^{(s-1)N/6}$ one can sum the binomial
series. The resulting expression is exponentially small in $N$ for any
$q<1$.

A lower bound on $Z_N$ can be calculated by neglecting configurations
with energy greater than $N/3-1$ as follows,
\begin{eqnarray}
Z_N & > & N
\sum_{m=0}^{N/3-2}q^m\sum_{m_i=0}^mP(m_1)P(m_2)P(m_3)\delta_{m_1+m_2+m_3,m}
\\ & = & N/[(q)_{\infty}]^3 - N
\sum_{m=N/3-1}^{\infty}q^m\sum_{m_i=0}^mP(m_1)P(m_2)P(m_3)\delta_{m_1+m_2+m_3,m}
\\ & > & N/[(q)_{\infty}]^3 - N \sum_{m=N/3-1}^{\infty}q^m (mP(m))^3 \; .
\label{lowbound}
\end{eqnarray}
The asymptotic behavior of $P(m)$ \cite{andrew} is given by
\begin{equation}
P(m) \simeq \frac{1}{4m \sqrt{3}} \exp{(\pi (2/3)^{1/2} \ m^{1/2})}\;.
\end{equation}
Thus, for large $N$ the lower bound (\ref{lowbound}) converges to
(\ref{psum}) as does the upper bound (\ref{zupperb1}).
\subsection{Correlation Functions}
Whether or not a system has long-range order in the steady state can
be found by studying the decay of two-point density correlation
functions. For example the probability of finding an $A$ particle at
site $i$ and a $B$ particle at site $j$ is,
\begin{equation}
\label{CorrF1}
\langle A_i B_j \rangle  = \frac{1}{Z_N} \sum_{\{X_k\}}
A_i B_j
 ~q^{{\cal H}(\{X_k\})} \; ,
\end{equation}
where the summation is over all configurations $\{X_k\}$ in which
$N_A=N_B=N_C$. Due to symmetry many of the correlation functions will
be the same, for example $\langle A_i A_j \rangle = \langle B_i B_j
\rangle =\langle C_i C_j \rangle$. A sufficient condition for
the existence of phase separation is
\begin{equation}
\label{CorrF2}
\lim_{r\rightarrow \infty}~\lim_{N\rightarrow \infty} (\langle A_1 A_r
\rangle - \langle A_1 \rangle \langle A_r \rangle) >0.
\end{equation} 
Since $\langle A_i \rangle =1/3$ we wish to show that
$\lim_{r\rightarrow \infty} ~\lim_{N\rightarrow \infty} \langle A_1
A_r \rangle > 1/9$.  In fact we will show below that for any given $r$
and for sufficiently large $N$,
\begin{equation}
\label{CorrF3}
\langle A_1 A_r \rangle = 1/3 - {\cal O}(r/N) \; .
\end{equation}
This result not only demonstrates that there is phase separation, but
also that each of the domains is pure. Namely the probability of
finding a particle a large distance inside a domain of particles of
another type is vanishingly small in the thermodynamic limit.

To prove Eq. (\ref{CorrF3}), we use the relation $\langle A_1A_r
\rangle = 1/3 - \langle A_1B_r \rangle - \langle A_1C_r \rangle$, and
show that the correlation function $\langle A_1B_r \rangle$ is of
${\cal O}(r/N)$ and $\langle A_1C_r \rangle$ is of ${\cal
O}(1/N)$. Here we show only the proof for $\langle A_1B_r \rangle$,
since the proof of $\langle A_1C_r \rangle$ is similar. We also
restrict ourselves to $r\le N/3$, which is sufficient for proving
Eq. (\ref{CorrF3}).

We have already seen that the contribution to the partition sum from
the metastable states and excitations above them are exponentially
small in the system size and hence may be neglected. Therefore, for
calculating the correlation function it is sufficient to consider the
$N$ ground states and excitations above them which may be reached by
moves which only increase the energy.  As we have seen, these states
form $N$ disjoint sets of states, each associated with one of the
ground states. Using this we now show that $\langle A_1B_r \rangle =
{\cal O}(r/N)$. For this purpose we use a restricted partition sum
${\cal Z}_s$, which is defined as the partition sum ${\cal Z}_N$
calculated with the constraint that one of the walkers, say of type
$A$, has traveled at least distance $s$.  It is given as $N \to
\infty$ by
\begin{equation}
\label{psums}
{\cal Z}_s = \sum_{m=0}^{\infty} q^m\sum_{m_i=0}^m P^s(m_1)P(m_2)P(m_3)
\delta_{m_1+m_2+m_3,m} \; .
\end{equation}
Here $P^s(m)$ is the number of partitions of integer $m$ with the
constraint that in all the partitions the integer $s$ occurs at least
once.  Noting that $P^s(m)=P(m-s)$ it is easy to show that
\begin{equation}
\label{psums1}
{\cal Z}_s = q^s {\cal Z} \; ,
\end{equation}
where ${\cal Z} \equiv \lim_{N \rightarrow \infty} {\cal Z}_N$.

We now proceed to derive a bound for $\langle A_1 B_r \rangle
$. Recall that $l$ is the position of the rightmost $A$ particle in
the $A$ domain in the ground state labeled $l$. If we define $\langle
A_1B_r \rangle_l$ as the correlation function calculated within the
set of states labeled $l$, we can write
\begin{equation}
\label{A1Br}
\langle A_1B_r \rangle = \frac{1}{N} \sum_{l=1}^{N} \langle A_1B_r
\rangle_l \; ,
\end{equation}
up to exponentially small corrections in the system size.  For
convenience we break the summation over $l$ into $4$ sums according to
the values of $A_1$ and $B_r$ in the ground state.  These $4$ parts
correspond to (I) ground states where $A_1=1, B_r=1$; (II) ground
states where $A_1=1, B_r=0$; (III) ground states where $A_1=0, B_r=1$
and (IV) ground states where $A_1=0, B_r=0$. We now consider each of
them in detail and give an upper bound for $\langle A_1 B_r \rangle_l$
in each case.

\begin{figure}
\centerline{\psfig{figure=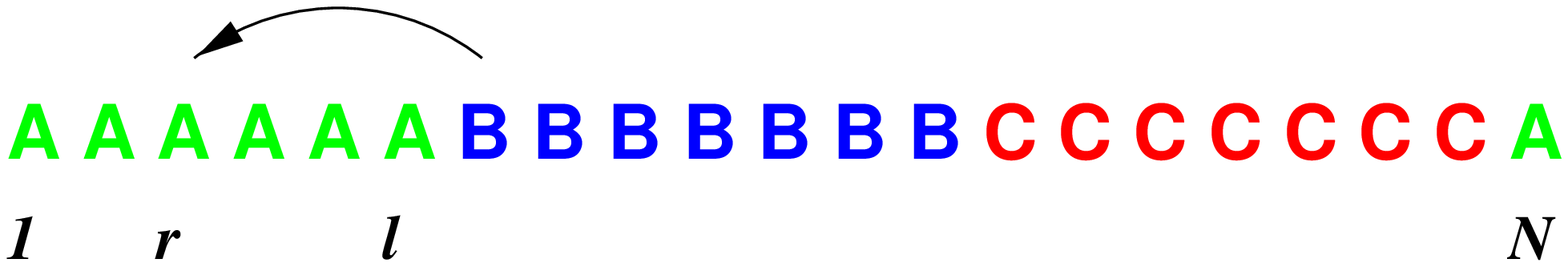,width=16cm}}
\caption{A ground state where $A_1=1,B_r=0$. In order to have $B_r=1$,
it is necessary for a 
$B$ particle to travel a distance of $l-r+1$ into the $A$
domain.}
\label{A10Bk1}
\end{figure}

\noindent(I) {\it Ground states where $A_1=1, B_r=1$}:
In this case the site $1$ is inside the $A$ domain and site $r$ is
inside the $B$ domain. Since we consider only $r\le N/3$, these states
correspond to the $l$ ground states with $1\le l < r$. Using the fact
that $\langle A_1 B_r \rangle_l \le 1$ one finds
\begin{equation}
\label{ubound1}
\sum_{l=1}^{r-1} \langle A_1 B_r \rangle_l \leq r-1\;.
\end{equation}
\noindent (II) {\it Ground states where $A_1=1, B_r=0$}:
in principle site $r$ might be either inside the $A$ domain or inside
the $C$ domain. However, since site $1$ is in the $A$ domain and we
consider only $r < N/3$, site $r$ must be in the $A$ domain.  The
ground states $l$ for which this takes place satisfy $r\le l \le
N/3$. Clearly, only the excited states where $B_r=1$ contribute to
$\langle A_1 B_r \rangle_l$.  In such excited states one of the $B$
walkers travels at least a distance $l-r+1$ into the $A$ domain (see
Fig. \ref{A10Bk1}).  For this case we can give the upper bound
$\langle A_1 B_r \rangle_l \le \sum_{s=l-r+1}^{\infty} {\cal
Z}_s/{\cal Z}$. From Eq. (\ref{psums1}), ${\cal Z}_s/{\cal
Z}=q^s$. Hence,
\begin{equation}
\label{ubound2}
\sum_{l=r}^{N/3} \langle A_1 B_r \rangle_l \leq \sum_{l=r}^{N/3} 
\sum_{s=l-r+1}^{\infty} q^s\;.
\end{equation}

\noindent(III) {\it Ground states where $A_1=0, B_r=1$}:
again, since $r\le N/3$ site $1$ has to be inside the $B$ domain.  The
values of $l$ satisfying this condition are in the range $2N/3+r \le l
\le N$.  In this case only excited states in which one of the $A$
walkers travels at least a distance $N-l+1$ into the $B$ domain will
contribute to $\langle A_1B_r \rangle_l$. Hence we can use the upper
bound $\langle A_1B_r \rangle_l \le \sum_{s=N-l+1}^{\infty} {\cal
Z}_s/{\cal Z}$, in this case. Therefore,
\begin{equation}
\label{ubound3}
\sum_{l=2N/3+r}^{N} \langle A_1 B_r \rangle_l \leq \sum_{l=2N/3+r}^{N} 
\sum_{s=N-l+1}^{\infty} q^s\;.
\end{equation}

\noindent(IV) {\it Ground states where $A_1=0, B_r=0$}: there are three
possibilities here. $(a)$ site $1$ is inside the $C$ domain and site
$r$ is inside the $A$ domain ($N/3 < l < N/3+r$), $(b)$ both the sites
$1$ and $r$ are inside the $C$ domain ($r+N/3 \le l \le 2N/3$), $(c)$
site $1$ inside the $B$ domain and site $r$ inside the $C$ domain
($2N/3 < l < 2N/3 +r$). Since all these are consistent with $r\le
N/3$, all these cases can occur. It can be shown that the minimal
energy needed to create an excited state where $A_1=1$ and $B_r=1$ is
$\epsilon_a=2l-r-N/3-1$ for the case $(a)$, $\epsilon_b=N/3+r-3$ for
the case $(b)$ and $\epsilon_c=5N/3-2l+r-1$ for the case $(c)$.  The
resulting expression for the bound is
\begin{equation}
\label{ubound4}
\sum_{l=N/3+1}^{2N/3+r-1}  \langle A_1 B_r \rangle_l
\leq \sum_{l=N/3+1}^{N/3+r-1}\sum_{s=\epsilon_a}^{\infty} q^s
+ \sum_{l=N/3+r}^{2N/3}\sum_{s=\epsilon_b}^{\infty} q^s
+ \sum_{l=2N/3+1}^{2N/3+r-1}\sum_{s=\epsilon_c}^{\infty} q^s \; .
\end{equation}
The summations on the RHS of Eqs. (\ref{ubound2}-\ref{ubound4})
can be carried out explicitly.  To leading order, the summations gives
$q/(1-q)^2$ for each of Eqs. (\ref{ubound2}) and
(\ref{ubound3}). The summation on the RHS of Eq. (\ref{ubound4})
vanishes exponentially in the thermodynamic limit. Using Eqs.
(\ref{A1Br}-\ref{ubound3}), we get the following expression for the
upper bound on $\langle A_1 B_r \rangle $
\begin{equation}
\langle A_1 B_r \rangle \leq \frac{1}{N} \left[r-1 + \frac{2q}{(1-q)^2} 
+e^{-{\cal O}(N)} \right] \;.
\end{equation}
Therefore $\langle A_1B_r \rangle={\cal O}(r/N)$.  Similarly one can
show that $\langle A_1C_r \rangle={\cal O}(1/N)$.  Thus for all $q<1$,
$\langle A_1A_r \rangle=1/3 - {\cal O}(r/N)$, proving the existence of
a complete phase separation.

\section{Coarsening}
\subsection{Monte Carlo Simulations}
We have demonstrated that in the thermodynamic limit the system is
phase separated when $N_A=N_B=N_C$. The general
arguments given in section II indicate that when the global densities
of the three species are non-vanishing and $q \neq 1$, the system
phase separates, even when the three densities are not equal. The
argument suggests that the typical time, $t_f$, in which the system
leaves a specific phase separated configuration increases
exponentially with the system size. Thus, a phase separated state is
stable in the thermodynamic limit. In the following we use Monte Carlo
simulations to support these arguments.

The time, $t_f$, can be measured using the auto-correlation function
defined as,
\begin{equation}
c(t)=\frac{1}{N} \sum_{i=1}^{N} ( \langle A_i(0)A_i(t) \rangle +
\langle B_i(0)B_i(t) \rangle + \langle C_i(0)C_i(t) \rangle ) \; ,
\end{equation}
where $A_i(t),B_i(t)$ and $C_i(t)$ are the values of the occupation
variables $A_i,B_i$ and $C_i$ at time $t$, and $\langle \ldots
\rangle$ denotes an average over histories of evolution. Clearly,
$c(0)=1$ while $c( \infty )=(N_A/N)^2+(N_B/N)^2+(N_C/N)^2$, the value
of the autocorrelation between two independent configurations. Thus,
$t_f$ may be defined as the decay time of $c(t)$ to $c( \infty )$ when
at $t=0$ the system is totally phase separated.

\begin{figure}[t]
\center{\psfig{figure=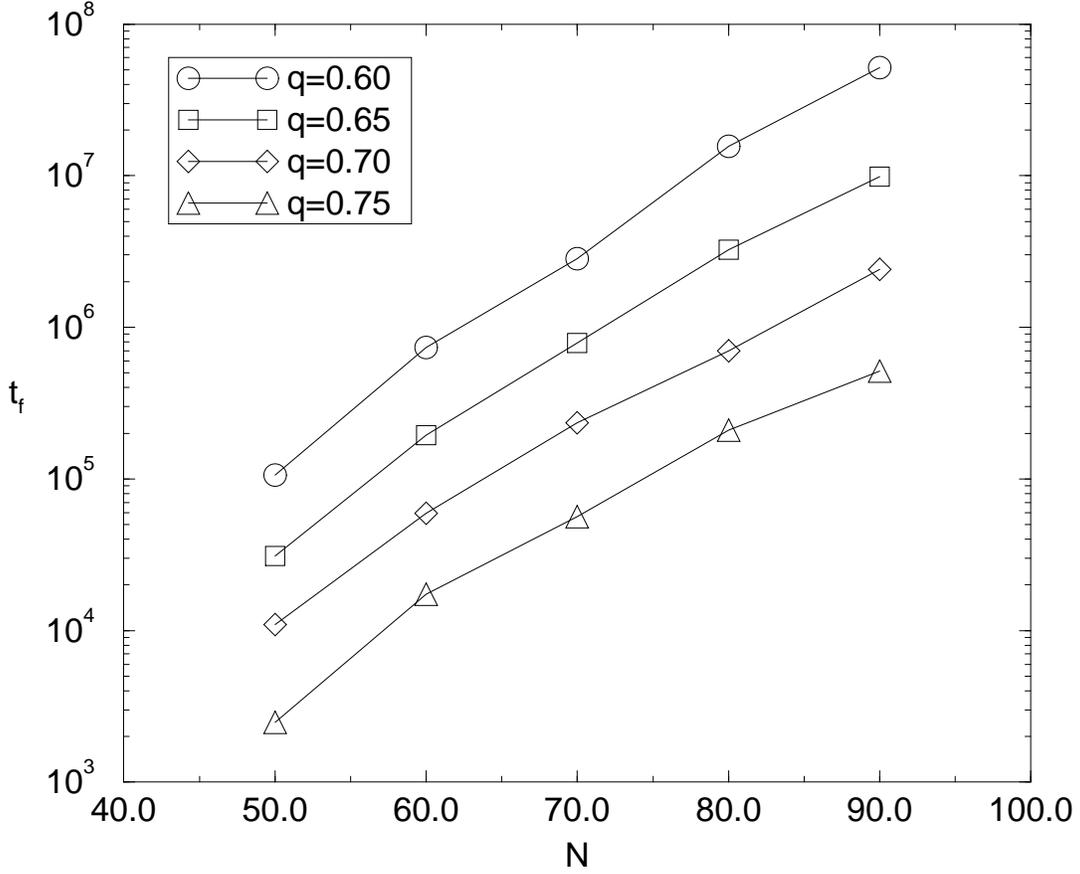,width=16cm}}
\caption{The decay time $t_f$ measured for different system sizes
 for several $q$ values. Here $N_A/N=0.4$, $N_B/N=0.35$ and
 $N_C/N=0.25$. The data is averaged over a 100 runs.}
\label{fig_corrt}
\end{figure}

We have measured the time scale $t_f$ using Monte Carlo simulations
for different system sizes for $N_A=N_B=N_C$ and for $N_A \neq N_B
\neq N_C$ for several $q$ values. An example of such  measurements for
$N_A/N=0.4$, $N_B/N=0.35$ and $N_C/N=0.25$ is presented in
Fig.~\ref{fig_corrt}. In the plot $t_f$ is plotted versus system size
for several values of $q$. It agrees with the exponential growth of
$t_f$ with the system size suggested by the simple argument of section
II. The same behavior seems to occur for all $q\ne1$ and different
choices of $N_A/N$, $N_B/N$ and $N_C/N$. Therefore we conclude that
the Monte Carlo simulations support the claim that for any $q \ne 1$
the system will phase separate into three domains in the thermodynamic
limit, even when the number of particles of each species is not
equal. In the thermodynamic limit the translational symmetry is
spontaneously broken in this state. Due to the slow dynamics, which
reflects escape from metastable states, Monte Carlo simulations could
be performed only for relatively small system size ($N \approx
100$). In order to study the coarsening process for larger systems we
employ, in the following, a toy model which mimics the dynamics of the
model (\ref{eq:dynamics}). The toy model can be conveniently simulated
for systems larger by about two orders of magnitude.

\subsection{Toy Model}
We now construct a simple toy model which captures the essential
physics of the coarsening process in the model at large times and
enables us to simulate systems much larger than those accessible by
Monte Carlo simulation. Using the toy model we examine another
characteristic scale of the system. Namely, the average domain size
$\langle l \rangle$ as a function of the time $t$. The results support
the simple argument leading to a domain growth law $\langle l \rangle
\sim \log t / \vert \log q \vert$. A mean-field version of the toy
model is then solved analytically.

\begin{figure}
\center{\psfig{figure=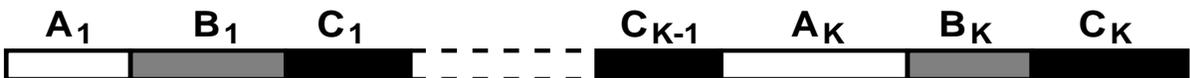,width=16cm}}
\caption{A configuration of the toy model represented by a sequence
 of domains.}
\label{fig_toy1}
\end{figure}

We consider a system at time $t$ such that the average domain size,
$\langle l \rangle$, is much larger than the domain wall width. At
these time scales, the domain walls can be taken as sharp and we may
consider only events which modify the size of domains. This means that
the dynamics of the system can be approximated by considering only the
movement of particles between neighboring domains of the same
species. Using this we represent a configuration by a sequence of
domains of the form ${\rm \bf A}_1{\rm \bf B}_1{\rm \bf C}_1{\rm \bf
A}_2{\rm \bf B}_2 {\rm \bf C}_2 \ldots {\rm \bf A}_K{\rm \bf B}_K{\rm
\bf C}_K$, where the $i$th domain of, say $A$, particles is
represented by ${\rm \bf A}_i$, as shown in Fig. \ref{fig_toy1}. The
exchange of particles between domains, say ${\rm {\bf A}}_i$ and ${\rm
{\bf A}}_{i+1}$, takes place at a rate dictated by the size of the
domains ${\rm {\bf B}}_i$ and ${\rm {\bf C}}_i$ which separate
them. Since intermediate configurations of the form $\ldots {\rm \bf
A}_{i-1} {\rm \bf B}_{i} {\rm \bf A}_{i} {\rm \bf C}_{i} \ldots $
rearrange on short time scales compared with the evolution between
metastable states, only metastable configurations are considered in
the toy model. Events in which a domain splits into two are ignored.

Using these ideas we define the dynamics of the toy model as follows: at each
time step two neighboring domains of the same species of particle are
chosen randomly, say ${\rm \bf A}_i$ and ${\rm \bf A}_{i+1}$. Let
$a_i$, $b_i$ and $c_i$ denote the lengths of the domains ${\rm \bf
A}_i,{\rm \bf B}_i$ and ${\rm \bf C}_i$ respectively. The length of
the domain chosen is then modified by carrying out one of the
following processes:
\begin{equation}
  \label{toy_rule}
  \begin{array}[]{ll}
    \left. 
     \begin{array}[]{ll}
    1) & a_i \rightarrow a_i - 1   \\
      & a_{i+1}\rightarrow a_{i+1} + 1  \\
     \end{array} \right\} & {\rm with~rate} \quad q^{b_i} \\
& \\
    \left.
     \begin{array}[]{ll}
    2) & a_i \rightarrow a_i + 1    \\
      & a_{i+1}\rightarrow a_{i+1} - 1  \\  
  \end{array} \right\} & {\rm with~rate} \quad q^{c_i} \\
\end{array}
\end{equation}
where, as before, $q < 1$ is considered.  

If $a_i$ becomes zero, then delete the domain ${\rm \bf A}_i$ from the
list of domains, and merge ${\rm \bf B}_i$ and ${\rm \bf C}_i$ with
${\rm \bf B}_{i-1}$ and ${\rm \bf C}_{i-1}$, respectively. Then for $j
> i$, shift the indices of the domains from $j$ to $j-1$, so that $K$
becomes $K-1$. The rules for updating ${\rm \bf B}$ and ${\rm \bf C}$
domains can be obtained from (\ref{toy_rule}) using cyclic
permutations and a slight change of indices.

Note that the toy model is only relevant to the description of the
coarsening dynamics. This is because here, once the system is left
with three domains, it remains in that state.

To simulate the toy model efficiently, an algorithm suitable for rare
event dynamics must be used due to the small rate of events
\cite{rare}. We use an algorithm which is performed by repeating the
following steps:

\begin{enumerate}
\item List all possible events $\{n\}$ and assign to them
rates $\{r_n\}$ according to the rules of the model.
\item Choose an event $m$ with probability  $r_m/R$ where $R=\sum_n r_n$.
\item Advance time by $t \rightarrow t + \tau $, where $\tau =1/r_m$  .
\end{enumerate}
The algorithm would be equivalent to a usual Monte Carlo simulation,
where $1$ time step is equivalent to one Monte Carlo sweep, if in step
$3$, $\tau$ would be drawn from a Poisson distribution $R ~\exp[-R
\tau ]$. However, here we make an approximation by using the
deterministic choice $ \tau =1/r_m$.

\begin{figure}[t]
\center{\psfig{figure=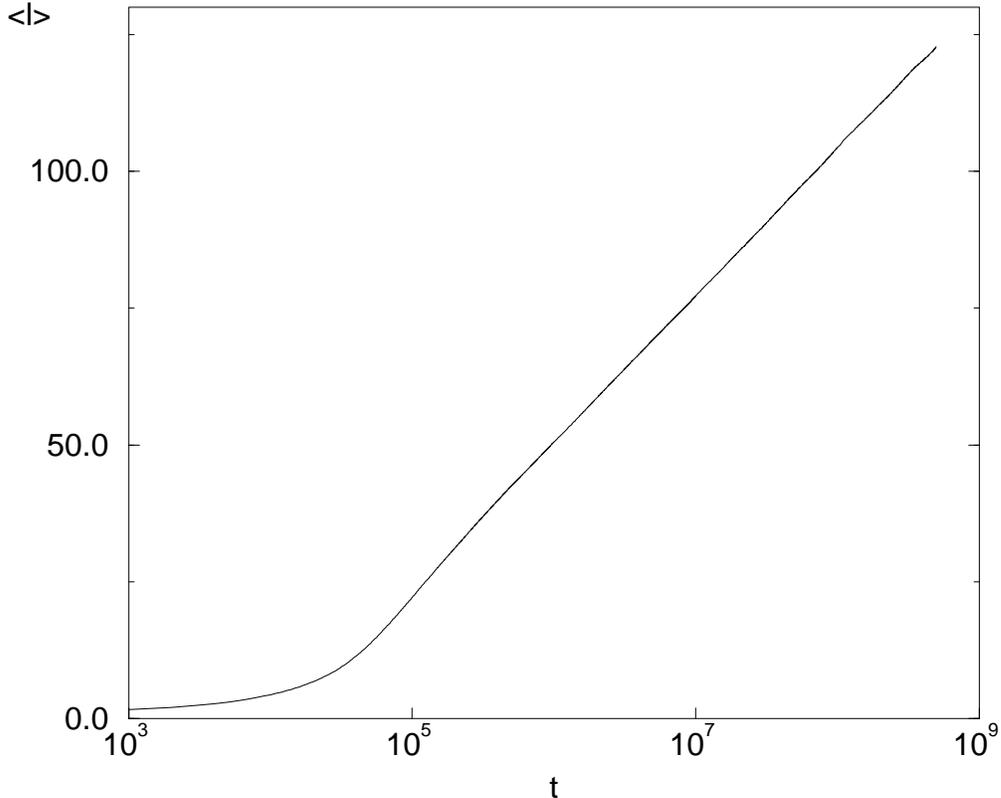,width=15cm}}
\caption{Monte Carlo simulation results for the toy model for the
 average domain size, $\langle l \rangle$, vs. time ,$t$, for 
$N=9000$ and $q=0.8$. The data is averaged over 1760 runs.}
\label{fig_domsize}
\end{figure}

We have simulated the dynamics for lattices of size up to $9000$. For
simplicity we consider the case $N_A=N_B=N_C$. An example of a typical
behavior of the average domain size as a function of $t$ is shown in
Fig. \ref{fig_domsize}. One can see that after an initial
transient growth time the data fits very well with a $\log(t)$
behavior. Simulations for different $q$ values indicate that,
\begin{equation}
\label{loga}
\langle l \rangle = a \log t/ \vert \log q \vert
\end{equation}
with $a \simeq 2.6$. The toy model enables one to verify the scaling
behavior (\ref{loga}) and estimate the constant $a$. This would be very
difficult to do by simulation of the full model
(\ref{eq:dynamics}).

\subsection {Mean Field Solution of the Toy Model}
\begin{figure}
\center{\psfig{figure=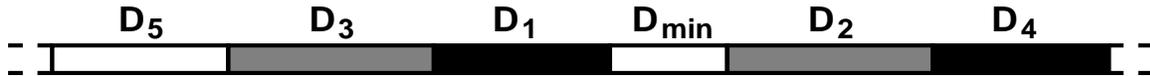,width=16cm}}
\caption{A configuration in the mean-field model after steps $1$ to $3$
 have been performed for one of the smallest domains.}
\label{fig_toy2}
\end{figure}

Here we present the solution of a mean-field version of the toy model
based on ideas presented by Rutenberg and Bray \cite{ARAB} and Derrida
{\it et al.}\cite{ajb_1} in the study of the ordering dynamics in a
one-dimensional scalar model. To construct the mean-field model we
notice that since all steps in the toy model which involve exchange of
particles between domains occur at a rate exponentially small in the
size of domains, one can consider a model where dynamics occurs only
in the vicinity of the smallest domains. The mean field approximation
assumes that
different domains are uncorrelated and does not distinguish between
domains of different species. The second assumption
relaxes the conservation of particles of each species. Thus, in
contrast to the systems studied by \cite{ARAB} and \cite{ajb_1} we do
not expect the mean field to become exact in the scaling limit. We
define the mean-field model as follows:

\begin{enumerate}
\item Pick one of the smallest domains $D_{\rm min}$. 
\item Pick $2$ domains $D_1$ and $D_2$ randomly and treat them
as the neighbors of $D_{\rm min}$.
\item Pick $3$ more domains randomly say, $D_3,D_4$ and $D_5$.
\item Eliminate $D_{\rm min},D_2$ and $D_3$ from the system
and change the length of the domains $D_3,D_4$ and $D_5$ by
\begin{eqnarray}
l_{D_3} & \rightarrow l_{D_3} + l_{D_2} \nonumber \\
l_{D_4} & \rightarrow l_{D_4} + l_{D_1} \label{toy2} \\
l_{D_5} & \rightarrow l_{D_5} + l_{D_{\rm min}} \; .\nonumber
\end{eqnarray}
\end{enumerate}
Steps $1$ to $4$ are performed simultaneously for all of the smallest
domains in the system. Here $l_{D_i}$ is the length of domain $i$, and
$l_{{\rm min}}$ is the length of the smallest domain. Steps $1$ to $3$
choose the smallest domain $D_{\rm min}$ and its nearest neighbors
(see Fig. \ref{fig_toy2}). Step $4$ uses the fact that the dynamics occur
in the model only in the vicinity of the smallest domain and
eliminates the three domains $D_{\rm min}$, $D_{1}$ and $D_{2}$
joining them appropriately with the other domains $D_{3}$, $D_{4}$ and
$D_{5}$. Note that this mean-field dynamics does not take into account
the time taken for these events to happen. This will be
done later when we derive the growth law of domains.

To solve the mean-field model we follow the method used in
\cite{ARAB,ajb_1}. Let $n_l(t)$ be the number of domains of size $l$,
irrespective of the type of particle it consists of. Let $l_{\rm
min}(t)$ be the length of the smallest domain and $M(t)=3K(t)$ be the
total number of domains at time $t$. We will assume that $n_l(t)$ has
the following scaling form in the large-time limit.
\begin{equation}
n_l = \frac{M}{l_{\rm min}} f(\frac{l}{l_{\rm min}})\;.
\end{equation}
A solution of the model given in Appendix \ref{meanfield} yields $\langle l
\rangle=\langle x \rangle l_{\rm min}$ where $\langle x \rangle$ is
given by
\begin{equation}
  \label{avexpp}
  \langle x \rangle = \frac{3 e^{\gamma/3}}{\int_0^{\infty}dx 
~ x^{-1/3}~e^{-x}~e^{-I(x)/3}} \; .
\end{equation}

The growth law for $l_{\rm min}(t)$ can be derived following
\cite{ARAB}: after the elimination of the smallest domain, $l_{\rm
min}$ increases by $1$. This happens at a rate $q^{l_{\rm min}}/
\langle l \rangle$, namely, the inverse time required by a typical
domain to cross a distance $l_{\rm min}$ (thus, causing the
annihilation of $D_{\rm min}$). Using $\langle l \rangle=\langle x
\rangle l_{\rm min}$ we write
\begin{equation}
\frac{\partial l_{\rm min}}{\partial t} = \frac{q^{l_{\rm min}}}
{\langle x \rangle l_{\rm min}} \; .
\end{equation}
From this one can obtain the scaling form of the average domain size,
\begin{equation}
\label{sform}
\frac{\langle l \rangle q^{-\langle l \rangle/\langle x \rangle}}{|\ln(q)|}
\approx t \; .
\end{equation}

Note that in this Eq. $\langle x \rangle$ does not depend,
according to the mean-field solution, on either $t$ or $q$. One can
see from (\ref{sform}) that for large $l$, $\langle l \rangle \approx
\langle x \rangle \ln t / \vert \ln q \vert$ which was confirmed by
the simulations of the toy model where $a=\langle x \rangle$ (see
(\ref{loga})). A numerical evaluation of (\ref{avexpp}) yields
$\langle x \rangle \simeq 3.72$ as compared with $a=2.6$ obtained from
the toy model simulations.
\section{Exact Results for Finite Systems}
\label{matrixans}
In section III the partition function, $Z_N$, and correlation
function, $\langle A_1A_r \rangle$ for finite $r$, have been
calculated in the thermodynamic limit. It is also of interest to
obtain results for finite systems for the study of finite-size effects
and the approach to the thermodynamic limit. Recently a matrix ansatz
method has been introduced to study one-dimensional non-equilibrium
systems \cite{DEHP}. It has been shown that in certain three-species models
 the
steady-state weight and correlation functions can be represented as a
product of matrices \cite{DJLS,EFGM1,Evans,AHR}. In the ansatz a specific matrix
is associated with each type of particle. Then the unnormalized
probability of a certain configuration is obtained from a matrix
product. The matrices corresponding to the different species of
particles satisfy an algebra derived from the dynamics of the model. A
scalar, {\it i.e.}  the weight of a configuration or some correlation
function, is usually obtained by performing a trace over the product
of matrices, or by multiplying both sides of the product of matrices
by vectors. Generalizing this method to replace matrices by tensors
\cite{DEM}, we have been able to obtain recursion relations for the
partition function and correlation function for finite systems for the
special case $N_A=N_B=N_C$. The recursion relations are then used to
obtain the partition function and correlation function $\langle A_1A_r
\rangle$ for any $r$ in small systems. The results are used to study
the scaling of the correlation function near the critical point $q=1$
(infinite temperature), where the typical domain wall width diverges.

\subsection {The Tensor Product Ansatz}
It is convenient to consider the unnormalized weights, $f_N(\lbrace
X_i \rbrace)$, defined through
\begin{equation}
\label{eq: weight2}
W_N(\lbrace X_i \rbrace)=Z_N^{-1}f_N(\lbrace X_i \rbrace) \; ,
\end{equation}
where $W_N(\lbrace X_i \rbrace)$ is the probability of being in
configuration $\lbrace X_i \rbrace$. The partition sum
$Z_N$ is given by
\begin{equation}
\label{eq: partition}
Z_N=\sum_{\lbrace X_i \rbrace}f_N(\lbrace X_i \rbrace) \; ,
\end{equation} 
where the sum is over all configurations with $N_A=N_B=N_C$.

We generalize the matrix ansatz and construct the steady-state weight,
 $f_N(\lbrace X_i \rbrace)$, from a product of
tensors each corresponding to a particle located in a specific place on
the lattice. The contraction of the tensors yields a tensor which is
then contracted with `left' and `right' tensors to generate
a scalar. The three tensors which represent the different type of
particle are defined as rank $6$ tensors through the following tensor
products of square matrices
\begin{eqnarray}
\mbox{\boldmath $\cal A$ \unboldmath}&=& {\bf E} \otimes {\bf D} \otimes {\bf 1} \nonumber \\
\mbox{\boldmath $\cal B$ \unboldmath}&=& {\bf 1} \otimes {\bf E} \otimes {\bf D}
\label{eq: tensors} \\
\mbox{\boldmath $\cal C$ \unboldmath} &=& {\bf D} \otimes {\bf 1} \otimes  {\bf E}\; . \nonumber
\end{eqnarray}
Here ${\bf 1}$ is a unit matrix. The matrices ${\bf D}$ and ${\bf E}$
will be chosen in what follows to satisfy a commutation relation which
will be dictated by the detailed balance condition.

To define $f_N(\lbrace X_i \rbrace)$ we introduce the following
notation: the contraction of two rank $6$ tensors $\mbox{\boldmath
${\cal O}$ \unboldmath}={\bf O_1}\otimes{\bf O_2}\otimes{\bf O_3}$ and
$\mbox{\boldmath ${\cal P}$ \unboldmath}={\bf P_1} \otimes {\bf P_2}
\otimes {\bf P_3}$, where ${\bf O_i}$ and ${\bf P_i}$ are square
matrices, according to the rule ${\bf O_1P_1}\otimes{\bf
O_2P_2}\otimes{\bf O_3P_3}$ is denoted by \boldmath ${\cal OP}$
\unboldmath. The contraction of a rank 6 tensor \boldmath ${\cal O}$
\unboldmath with a `left' rank $3$ tensor $\langle {\cal K} \vert =
\langle K_1 \vert \otimes \langle K_2 \vert \otimes \langle K_3
\vert$, where $\langle K_i \vert$ are transposed vectors, and a
`right' rank $3$ tensor, $\vert {\cal M} \rangle = \vert M_1 \rangle
\otimes \vert M_2 \rangle \otimes \vert M_3 \rangle$, where $\vert M_i
\rangle$ are vectors, defined through $\langle K_1 \vert {\bf O_1}
\vert M_1 \rangle \langle K_2 \vert {\bf O_2} \vert M_2 \rangle
\langle K_3 \vert {\bf O_3} \vert M_3 \rangle$ is denoted by $\langle
{\cal K} \vert \mbox{ \boldmath ${\cal O}$ \unboldmath} \vert {\cal M}
\rangle$.

Using these definitions we write the steady-state weight of the system
as
\begin{equation}
\label{eq: ansatz}
f_N(\lbrace X_i \rbrace)=\langle {\cal U}
 \vert \prod_{i=1}^N
[A_i\mbox{\boldmath $\cal A$ \unboldmath}
+B_i\mbox{\boldmath $\cal B$ \unboldmath}
+C_i\mbox{\boldmath $\cal C$ \unboldmath}] \vert {\cal V} \rangle \; ,
\end{equation}
where $A_i$, $B_i$ and $C_i$ are the occupation variables defined in
(\ref{ocup}).  The expression states that a tensor \boldmath $\cal A$
\unboldmath is present at place $i$ in the tensor product if site $i$
is occupied by an $A$ particle, a tensor \boldmath $\cal B$
\unboldmath is present if site $i$ is occupied by a $B$ particle, and
a tensor \boldmath $\cal C$ \unboldmath is present if site $i$ is
occupied by a $C$ particle. The action of the tensor product on
$\langle {\cal U} \vert $ and $\vert {\cal V} \rangle$ (to be
specified later) produces the scalar $f_N(\lbrace X_i \rbrace)$.

It is straightforward to show using detailed balance that a necessary
condition for Eq. (\ref{eq: ansatz}) to be the steady-state
weight is that the following commutation relations are satisfied
between \boldmath ${\cal A}$, ${\cal B}$, and ${\cal C}$ \unboldmath:
\begin{eqnarray}
q\mbox{ \boldmath $ {\cal A}{\cal B} $\unboldmath}
&=&\mbox{ \boldmath $ {\cal B}{\cal A}$\unboldmath }\nonumber \\
q\mbox{ \boldmath ${\cal B}{\cal C} $\unboldmath}&=&
\mbox{ \boldmath $ {\cal C}{\cal B}$ \unboldmath }
\label{eq: commutation}\\
q\mbox{ \boldmath ${\cal C}{\cal A} $\unboldmath}&=&
\mbox{ \boldmath $ {\cal A}{\cal C}$ \unboldmath }\; .\nonumber
\end{eqnarray}
Using (\ref{eq: tensors}) one can verify that these commutation
relations are satisfied provided that $q{\bf DE}={\bf ED}$. This
deformed commutator is of relevance in other
stochastic systems \cite{MAG,SchS}. A representation of the matrices
${\bf D}$ and ${\bf E}$ which satisfies this commutation relation can
be obtained as follows: let $\lbrace \langle n \vert \rbrace$ denote a
basis set ($n=0,1,\ldots,N/3$) forming a vector space. In this basis
we choose the matrices so that,
\begin{equation}
\label{eq: matrix}
\begin{tabular}{l}
$\langle n \vert {\bf E}= \langle n \vert q^n\;\;\;\;\;$ for any $n$\\
$\langle n \vert {\bf D}= \langle n-1 \vert \;\;$ for $n \geq 1$,\\ 
\end{tabular}
\end{equation}
while for $n=0$, $\langle 0 \vert {\bf D}=0$. An explicit form for
${\bf E}$ and ${\bf D}$ is given by the following $(N/3+1) \times
(N/3+1)$ square matrices
\begin{equation}
\label{eq: matrixDE}
{\bf E} = \sum_{n=0}^{N/3} q^n \vert n \rangle
\langle n \vert \;\;\; ; \;\;\;
{\bf D} = \sum_{n=1}^{N/3}  \vert n \rangle
\langle n-1 \vert \; .
\end{equation}

To obtain $f_N(\lbrace X_i \rbrace)$, $\langle {\cal U} \vert$ and
$\vert {\cal V} \rangle$ have to be specified. This should obviously
be done so that $f_N(\lbrace X_i \rbrace)$ is non-zero if the ansatz
is to give a non-trivial result. We consider a general tensor product
which corresponds to some configuration. The product has $N/3$ tensors
of each type \boldmath ${\cal A},{\cal B}$ and ${\cal C}$ \unboldmath,
which results in a tensor product of three matrix products. Using
(\ref{eq: tensors}) it can be seen that each matrix product contains
$N/3$ matrices of each type ${\bf D}$, ${\bf E}$ and ${\bf 1}$. Since
${\bf E}$ and ${\bf 1}$ are diagonal, while ${\bf D}$ acts to its left
as a lowering matrix (see (\ref{eq: matrix})), choosing $\langle {\cal
U} \vert=\langle N/3 \vert \otimes \langle N/3 \vert \otimes \langle
N/3 \vert \equiv \langle N/3,N/3,N/3 \vert$, and $\vert {\cal V}
\rangle = \vert 0 ,0,0 \rangle$ will give a non-zero $f_N(\lbrace X_i
\rbrace)$. This makes clear that the minimal size choice for the
vector-spaces is $N/3+1$. Under this choice it is easy to see that in
the ground states one has $f_N=q^{N^2/9}$ which corresponds to a
ground state energy $N^2/9$.  However, the choice of $\langle {\cal U}
\vert$ and $\langle {\cal V} \vert$ is determined only up to some
multiplicative factors. These factors may be used to shift the ground
state energy of the system. For example choosing $\langle {\cal U}
\vert$ as before with $\vert {\cal V} \rangle = q^{-N^2/9} \vert 0
,0,0 \rangle$ will shift the ground state energy to $0$. In the
following the factors are taken to be $1$.

Finally we would like to remark that usually when using the matrix
ansatz for systems with periodic boundary conditions it is often
convenient to use a trace of the matrix product \cite{DJLS,AHR}. In
this case this is not possible since the trace of our tensor product
is always zero.

\subsection {Partition Sum}
In terms of the ansatz (\ref{eq: ansatz}) the partition function $Z_N$
is given by
\begin{equation}
\label{eq: matrix partition}
Z_N=\langle \frac{N}{3},\frac{N}{3},\frac{N}{3}
 \vert (\mbox{\boldmath $ {\cal A}+{\cal B}+{\cal C}$\unboldmath })^N \vert
0,0,0 \rangle \; .
\end{equation}
Note that we are using the canonical ensemble since all tensor
products with unequal number of particles do not contribute to
$Z_N$. This is easily seen since in these cases there are always more
than $N/3$ ${\bf D}$ matrices acting on one of the vectors $\langle
N/3 \vert$. 

To obtain the partition function we derive a recursion relation for
\begin{equation}
G_{i,j,k}^l \equiv \langle i, j, k \vert (\mbox{\boldmath ${\cal
A}+{\cal B}+{\cal C}$\unboldmath })^l \vert 0, 0, 0 \rangle\;.
\end{equation}
One can
see that $G_{N/3,N/3,N/3}^N=Z_N$. Rewriting $G_{i,j,k}^l$ as
\begin{eqnarray}
\label{eq: prerecurr}
G_{i,j,k}^l & = & \langle i, j, k \vert 
\mbox{ \boldmath $ {\cal A}({\cal A}+{\cal B}+{\cal C})
$\unboldmath}^{l-1} \vert 0, 0, 0 \rangle + \langle i, j, k \vert
\mbox{ \boldmath $ {\cal B}({\cal A}+{\cal B}+{\cal C}
$ \unboldmath})^{l-1} \vert 0, 0, 0 \rangle \nonumber
\\ & & + \langle i, j, k \vert
\mbox{ \boldmath ${\cal C}({\cal A}+{\cal B}+{\cal C}
$ \unboldmath})^{l-1} \vert 0, 0, 0 \rangle \; ,
\end{eqnarray}
and using relations (\ref{eq: tensors}) and (\ref{eq: matrix}) the
following recursion relation can be derived
\begin{equation}
\label{eq: reccursion1}
G_{i,j,k}^l=q^iG_{i,j-1,k}^{l-1}+q^jG_{i,j,k-1}^{l-1}+q^kG_{i-1,j,k}^{l-1} \; . 
\end{equation}
The boundary conditions for this recursion relation is given by the no
particle partition function $G_{i,j,k}^0=1$ if $i=j=k=0$
and is zero otherwise.

For small systems (up to $N=21$) for which the recursion relation is
tractable analytically on Mathematica, we obtained the partition
function $Z_N=G_{N/3,N/3,N/3}^N$ as a polynomial in $q$. As expected
the first $N/3-2$ terms of the polynomial match the first $N/3-2$
terms of the expansion of (\ref{psum}) up to a factor of $q^{N^2/9}$
due to the energy shift in the ground state. For larger $N$ we solve
the recursion relation numerically.

We note that (\ref{eq: reccursion1}) could have been derived directly
from the definition of the partition function without recourse to the
tensor ansatz. However, we believe the utility of the ansatz lies in
the ease with which correlation functions can be manipulated and
relations such as that of the next subsection derived.

\subsection {Correlation Functions}
The correlation function $\langle A_1A_r \rangle$ is given in terms of
the ansatz by
\begin{equation}
\label{eq: corr1}
\langle A_1A_r \rangle =\frac{ \langle \frac{N}{3},\frac{N}{3},\frac{N}{3}
\vert 
\mbox{ \boldmath ${\cal A}({\cal A}+{\cal B}+{\cal C})$ \unboldmath }^{r-2}
\mbox{ \boldmath ${\cal A}({\cal A}+{\cal B}+{\cal C})$ \unboldmath }^{N-r}
\vert 0, 0, 0 \rangle}{Z_N} \; .
\end{equation}   
Using relations (\ref{eq: tensors}),(\ref{eq: commutation}) and (\ref{eq:
matrix}) we obtain
\begin{eqnarray}
\langle A_1A_r \rangle & = & q^{\frac{2N}{3}}\frac{ \langle
\frac{N}{3}, \frac{N}{3}-2,
\frac{N}{3}\vert 
(\mbox{\boldmath ${\cal A}$ \unboldmath }
+q \mbox{\boldmath ${\cal B} $ \unboldmath }
+\mbox{\boldmath ${\cal C}$ \unboldmath }/q )^{r-2}
(\mbox{\boldmath $ {\cal A}+{\cal B}+{\cal C}$ \unboldmath }  )^{N-r}
 \vert 0 , 0 , 0 \rangle}{Z_N}\\ & = &
q^{\frac{2N}{3}}\frac{U(r)^r_{\frac{N}{3},\frac{N}{3}-2,\frac{N}{3}}}{Z_N} \; ,
\label{A1Ar}
\end{eqnarray}
where we define the object $U(r)^{s}_{i,j,k}$ through,
\begin{equation}
\label{eq: U}
U(r)^{s}_{i,j,k}=\langle i, j, k \vert
(\mbox{\boldmath ${\cal A}$ \unboldmath }
+q \mbox{\boldmath ${\cal B} $ \unboldmath }
+\mbox{\boldmath ${\cal C}$ \unboldmath }/q )^{s-2}
(\mbox{\boldmath $ {\cal A}+{\cal B}+{\cal C}$ \unboldmath }  )^{N-s}
 \vert
0, 0, 0 \rangle \; .
\end{equation}
A recursion relation for $U(r)^{s}_{i,j,k}$ can be
obtained, similarly to the recursion relation for the partition
function (\ref{eq: reccursion1}). Using (\ref{eq: tensors}) and (\ref{eq:
matrix}) gives
\begin{equation}
\label{eq: U recurr}
U(r)^{s}_{i,j,k}=q^iU(r)^{s-1}_{i,j-1,k}+q^{j+1}
U(r)^{s-1}_{i,j,k-1}+q^{k-1}U(r)^{s-1}_{i-1,j,k} \; .
\end{equation}
The boundary conditions for the recursion relation are obtained by
noting that $U(r)^2_{i,j,k}=\langle i, j, k \vert 
\mbox{ \boldmath $({\cal A}+{\cal B}+{\cal C})
$\unboldmath }^{N-r} \vert 0, 0, 0 \rangle$, {\it i.e.},
$U(r)^2_{i,j,k}=G_{i,j,k}^{N-r}$. Using the same methods one can
obtain recursion relations for all other correlation functions.

The recursion relations are solved numerically for finite
systems. This is done by first solving numerically for
$G_{i,j,k}^{N-r}$, and then using the result as boundary conditions
for the recursion relation (\ref{eq: U recurr}).  Owing to
(\ref{A1Ar}) we are ultimately interested in $U(r)^r_{N/3,N/3-2,N/3}$.
\begin{figure}[t]
\center{\psfig{figure=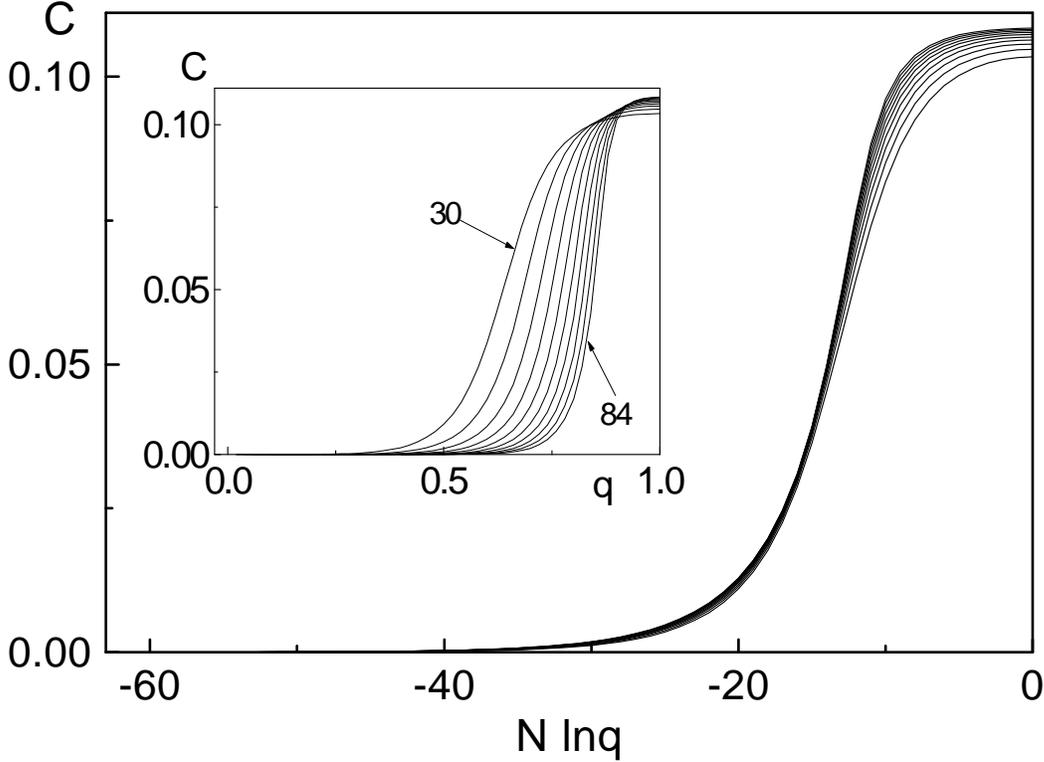,width=16cm}} 
\caption{The correlation function $C=\langle A_1A_{N/2} \rangle$, obtained from the tensor ansatz, as a function of the scaled variable $N \ln q$ for
$N=30,36,42, \ldots ,84$. The inset shows the same data plotted
against $q$.} 
\label{order}
\end{figure}

Using these recursion relations we have calculated the correlation
function $\langle A_1A_{N/2} \rangle$, which is a measure of the phase
separation in the system, for various system sizes. When $\langle
A_1A_{N/2} \rangle$ is close to zero the system is phase separated. In
the disordered case, $q=1$, the value of the correlation function is
$(N-3)/9(N-1)$, approaching $1/9$ in the thermodynamic limit. The
results are shown in the inset of Fig. \ref{order} one can see that
the system is phase separated for small values of $q$ while for $q$
close to $1$ the system is disordered. The range of $q$ values for
which the system is phase separated increases as the system size
increases. The natural scaling variable near the critical point $q=1$
is $N \ln q$, the ratio between the domain size $N/3$ and the domain
wall width $\int l q^l dl / \int q^l dl = 1/\vert \ln q \vert$. In
Fig.(\ref{order}) the correlation function is plotted as a function of
the scaling variable. One can see that the data collapse improves
as the system size increases.
This scaling variable was also suggested
by the form of the partition function (\ref{psum})
in section III C.

\section{Generalization to $M$ species}
In the following we discuss possible generalization of the model to $M
\geq 3$ species. To demonstrate how this might be done we first discuss
the case $M=4$. We then comment briefly on $M>4$.

 We now define a $4$-species model and argue that it phase
separates. Consider a ring where each site is occupied by either an
$A$, $B$, $C$ or $D$ particle. The model evolves according to the
following procedure: at each time step two nearest neighbors are
chosen randomly and exchanged according to the rates,

\begin{equation}
\label{eq:four}
\begin{picture}(130,80)(0,2)
\unitlength=1.0pt
\put(36,75){$AB$}
\put(56,73) {$\longleftarrow$}
\put(62,69) {\footnotesize $1$}
\put(56,77) {$\longrightarrow$}
\put(62,82) {\footnotesize $q$}
\put(80,75){$BA$}
\put(36,55){$BC$}
\put(56,53) {$\longleftarrow$}
\put(62,49) {\footnotesize $1$}
\put(56,57) {$\longrightarrow$}
\put(62,62) {\footnotesize $q$}
\put(80,55){$CB$}
\put(36,35){$CD$}
\put(56,33) {$\longleftarrow$}
\put(62,29) {\footnotesize $1$}
\put(56,37) {$\longrightarrow$}
\put(62,42) {\footnotesize $q$}
\put(80,35){$DC$}
\put(36,15){$DA$}
\put(56,13) {$\longleftarrow$}
\put(62,09) {\footnotesize $1$}
\put(56,17) {$\longrightarrow$}
\put(62,22) {\footnotesize $q$}
\put(80,15){$AD$}
\put(36,-5){$AC$}
\put(56,-7) {$\longleftarrow$}
\put(62,-11) {\footnotesize $1$}
\put(56,-3) {$\longrightarrow$}
\put(62,2) {\footnotesize $q$}
\put(80,-5){$CA$}
\put(36,-25){$DB$}
\put(56,-27) {$\longleftarrow$}
\put(62,-31) {\footnotesize $1$}
\put(56,-23) {$\longrightarrow$}
\put(62,-18) {\footnotesize $q$}
\put(80,-25){$BD$}.
\end{picture}
\end{equation}
\vspace{0.6cm}

\noindent As before the model conserves the number of particles of each
species. Note that several other generalizations of the model to four
species are possible. However, for simplicity, we discuss only the
model defined by (\ref{eq:four}) with $q<1$.

We now argue that the system phase separates into a configuration of
the form $ABCD$ (where each letter now indicates a domain) as long as
the densities of particles of each species are non-zero. Note that in
the model $AB$, $BC$, $CD$, $DA$, $AC$ and $DB$ boundaries are stable
while reverse boundaries $BA$, $CB$, $DC$, $AD$, $CA$ and $BD$ are
unstable. As in the case of the three species model the system,
starting from a random initial condition, evolves on a short time
scale ({\it i.e.}, which is not determined by the size of the system)
into a metastable configuration where only stable domains are present.
This configuration then slowly coarsens by slow diffusion of particles
through neighboring domains. The system will finally reach the most
stable state where the number of domains is minimal. One can easily
check that this configuration is given by $ABCD$. Note that the system
may exhibit other metastable states. For example, a state composed of
$ACDABCD$ is also stable under the choice of rates
(\ref{eq:four}). However, since this state is composed of more domains
than the $4$-domain state, some of the domains are necessarily
smaller. According to the argument presented in Section II the
relaxation time of this sequence (proportional to $q^{-m}$ where $m$
is the typical domain size) is much shorter than the relaxation time
of the $4$-domain state. Therefore the $4$-domain state is more stable
so that the system will finally evolve into it.

In considering $M>4$ models one finds that for some choices of
transition rates several states with the minimum number of domains may
become metastable. For example, for $M=5$ it is possible to choose
transition rates for which both $ABCDE$ and $ACEBD$ are locally
stable. The relative stability (and thus the resulting phase separated
state) may be found by determining the relaxation time of these states
using simple considerations such as those presented in Section II.

As is the case of $M=3$ detailed balance is found to be satisfied for
certain densities and transition rates for $M>3$. The condition for
this is derived in Appendix \ref{dbappen}. In this case the
relative stability of metastable states could be determined by
comparing free energies.

\section{Conclusion}
In this paper a model of three species of particles diffusing on a
ring previously introduced in \cite{EKKM} has been studied. The model
is governed by local dynamics in which all moves compatible with
the conservation of the three densities are allowed. We argue that
phase separation should occur as long as all densities are
non-zero. In the special case of equal densities we find that the
steady state generated by the local stochastic dynamics is exactly
given by a long-range asymmetric Hamiltonian. Phase separation for
this case is explicitly demonstrated. The model provides an explicit
example for the mechanism leading to phase separation or breaking of
ergodicity in systems with {\it local} stochastic dynamics. Although
we did not succeed in solving the steady state in the case of
non-equal densities of particles, there is a strong evidence that
phase separation still occurs. In order to investigate further the
case of equal densities we employed a generalized matrix ansatz to
calculate correlation function for finite-size systems. The novel
structure of the ansatz may give some clue as to handle other $1d$
models which have so far resisted solution.

The dynamics of phase separation reduces to a coarsening problem where
the typical domain size grows logarithmically in time. This results
from the elimination of the domains at a rate exponentially small in
their size. The slow dynamics poses a problem of how to access the
scaling regime numerically. With direct numerical simulations only
small systems can be studied (see Fig.~3). However, by employing a toy
model in which domains rather than individual sites are updated one
can simulate much larger systems and probe the scaling regime (see
Fig.~5). Such ideas of updating domains have been used before in the
study of coarsening \cite{CB}. With the aid of the toy model it should
be possible to study other aspects of the scaling regime associated
with the slow dynamics and escape from metastable states.

Generalizations are possible to models with $M>3$ species. We have
discussed some possibilities and have shown that phase separation may
take place, although the structure of the set of metastable states is
more complicated. As was the case for $M=3$, conditions for detailed
balance with respect to a long-range Hamiltonian may be determined.

The problem of phase separation and coarsening is of interest also in
the broader context of phase transitions in one dimensional
systems. Here the existence of conserved quantities results in certain
local transition rates being zero. It would be interesting to
generalize this study to models in which no conserved quantity exists
and all local rates are non-vanishing. Also, another open problem is
to calculate the steady state of the present model in the case of
non-equal densities.

\vspace{0.2cm}
\noindent {\bf Acknowledgments}: MRE is a Royal Society University
 Research Fellow and thanks the Weizmann Institute and the Einstein
Center for warm hospitality during several visits. The support of the
Minerva Foundation, Munich Germany, the Israeli Science Foundation, and
the Israel Ministry of Science is gratefully
acknowledged. Computations were performed on the SP2 at the
Inter-University High Performance Computer Center, Tel Aviv. We thank
S. Wiseman for helpful advice on programming and M. J. E. Richardson
for careful reading of the manuscript.

\appendix 
\section{Mean-Field Solution of Toy Model}
\label{meanfield}
To solve the mean-field model we follow the method used in
\cite{ARAB,ajb_1}. Let $n_l(t)$ be the number of domains of size $l$,
irrespective of the type of particle it consists of. Let $l_{\rm
min}(t)$ be the length of the smallest domain and $M(t)=3K(t)$ be the
total number of domains at time $t$. We will assume that $n_l(t)$ has
the following scaling form in the large-time limit.
\begin{equation}
n_l = \frac{M}{l_{\rm min}} f(\frac{l}{l_{\rm min}})
\label{nlform}
\end{equation}
After the elimination of the smallest domain as given by
(\ref{toy2}), $n_l,l_{\rm min}$ and $M$ change according to
\begin{eqnarray}
\label{arr}
  M' &=& M - 3 n_{l_{\rm min}} \\
  n'_l&=&n_l(1 - 5 \frac{n_{l_{\rm min}}}{M}) + n_{l_{\rm min}} 
 \frac{n_{l-l_{\rm min}}}{M} \theta (l-2 l_{min}) +
 2 n_{l_{\rm min}} \sum_{j=l_{\rm min}}^{l-l_{\rm min}} \frac{n_j}{M}
  \frac{n_{l-j}}{M} \label{modif} \\
  l'_{\rm min}&=& l_{\rm min} + 1 \; . 
\end{eqnarray}
Using the scaling form (\ref{nlform})  we have,
\begin{equation}
  \label{expan}
  n'_l = \frac{M'}{l_{\rm min} + 1} f(\frac{l}{l_{\rm min}+1})
  \approx \frac{M}{l_{\rm min}} [f(x) - (3f(1) + 1) \frac{f(x)}{l_{\rm
  min}} - \frac{x}{l_{\rm min}} \partial_x f(x)]
\end{equation}
where $x=l/l_{\rm min}$ and we have expanded in $1/l_{\rm min}$.
Now substituting these in (\ref{modif}), it is straightforward to show that
\begin{equation}
\label{eqf}
f(x) + x \partial_x f(x) - 2 f(x) f(1) + f(1) f(x-1) \theta (x-2)
+ 2 \theta (x-2) f(1) \int_{1}^{\infty} dy ~~ f(y)f(x-y) = 0 \; .
\end{equation}
Using the Laplace transform,
\begin{equation}
  \label{laplace}
  \phi(p) = \int_1^{\infty}dx ~~\exp[-px]~f(x)
\end{equation}
one can show that $\phi(p)$ satisfies the differential equation,
\begin{equation}
  \label{phip}
  p \partial_p \phi(p) = f(1)[\phi(p) - 1][2 \phi(p) + e^{-p}]
\end{equation}
Since $\phi(p) = 1 - \langle x \rangle p + \ldots$, by expanding
(\ref{phip}) to order $p$ one obtains $f(1) = 1/3$.
The solution of (\ref{phip}) with boundary conditions $\phi(0) = 1$
and $\phi(p) \approx e^{-p}/3p$ for $p>>1$ is
\begin{equation}
  \label{solu}
  \phi(p) = \frac{\int_p^{\infty}dx ~~ x^{-1/3}~e^{-x}~e^{-I(x)/3}}
{3p^{2/3}e^{-I(p)/3} + \int_p^{\infty}dx ~~ x^{-1/3}~e^{-x}~e^{-I(x)/3}}
\end{equation}
where $I(x) = \int_1^{\infty}dt ~~\exp[-xt]/t = -\log(x) - \gamma -
\sum_{n=1}^{\infty} (-x)^n/(n!\ n)$ and $\gamma$ is the Euler
constant.  Inverse Laplace transform of (\ref{solu}) gives the domain
size distribution. From (\ref{solu}) and using the expansion $\phi(p)
= 1 - \langle x \rangle p + \ldots$, where the average is with respect
to $f(x)$, we get
\begin{equation}
  \label{avex}
  \langle x \rangle = \frac{3 e^{\gamma/3}}{\int_0^{\infty}dx 
~ x^{-1/3}~e^{-x}~e^{-I(x)/3}}\;.
\end{equation}
\section{Detailed Balance Condition for an M-Species Model}
\label{dbappen}
We now define the most general $M$ species model, where $M \ge 3$. Let
$X_i$ denote a variable at site $i$ of a ring of size $N$, which
takes values $X_i=1,2,\ldots,M$. $X_i=m$ means that site $i$ is
occupied by a particle of type $m$. The system evolves by a random
sequential, nearest-neighbor exchange dynamics, with the following
rates:

\begin{equation}
\label{eq:Mspeci}
mn ~~{{q(m,n) \atop \longrightarrow}\atop {\longleftarrow \atop q(n,m)}} ~~nm \; ,
\end{equation}
and $q(X_i,X_i)=1$. The model conserves $N_m$, the number of particles
of type $m$, for all $m$.

We now present a condition for the model to satisfy detailed balance
with respect to the steady-state weight given by
\begin{equation}
  \label{eq:sweight}
  W(\{X_i\}) = \prod_{i=1}^{N-1} \prod_{j=i+1}^N q(X_j,X_i) \; ,
\end{equation}
where the set $\{X_i\}$ describes the microscopic configuration.

Consider a particle exchange between sites $k$ and $k+1$, where
$X_k=m,X_{k+1}=n$ and $k\ne N$ ({\it i.e.} in the bulk, note
that site $1$ is chosen arbitrarily). 
Expanding the product in (\ref{eq:sweight}), it is easy to verify that
\begin{equation}
  \frac{W(X_1,\ldots,m,n,\ldots,X_N)}{W(X_1,\ldots,n,m,\ldots,X_N)}  =
 \frac{q(n,m)}{q(m,n)} \; .
\end{equation}
Since this hold for any $m,n$, and is irrespective of the number of
particles of each species, the steady-state weight (\ref{eq:sweight})
satisfies detailed balance for all nearest-neighbor exchanges in the
bulk. If the weights (\ref{eq:sweight}) are translationally invariant
then detailed balance will also hold for exchanges between sites $1$
and $N$.

Thus, to complete the proof of detailed balance it is sufficient to
demand that (\ref{eq:sweight}) is translationally invariant. To do
this we relabel sites $i \rightarrow i+1$. The weight then becomes
\begin{equation}
 \label{eq:sweight2}
  W(\{X_i\}) = \prod_{i=1}^{N-1} \prod_{j=i+1}^N q(X_{j-1},X_{i-1}) \; ,
\end{equation}
where $X_0$ is identical to $X_N$.
Rewriting this equation by relabeling the indices we obtain,
\begin{equation}
 \label{eq:sweight3}
  W(\{X_i\}) = \left[\ \prod_{i=1}^{N-1} \prod_{j=i+1}^{N} q(X_{j},X_{i})\
 \right] \
  \prod_{k=1}^{N-1}\frac{q(X_k,X_N)}{q(X_N,X_k)} \; .
\end{equation}
Comparing (\ref{eq:sweight3}) with (\ref{eq:sweight}) and noting for
example that,
\begin{equation}
\prod_{j=1}^{N}q(X_j,X_N) = \prod_{l=1}^{M} \left[ q(l,X_N) \right]^{N_l} \; ,
\end{equation}
one can see that (\ref{eq:sweight}) is translational invariant if
\begin{equation}
\label{condition}
 \prod_{l = 1}^M \left[\frac{q(m,l)}{q(l,m)}
                     \right]^{N_{l}} = 1 \; , 
\end{equation}
for every $m=1,\ldots,M$. Thus, detailed balance holds if
(\ref{condition}) is satisfied. We note that for given densities
$\{ N_m \}$ the manifold of solutions for the rates is of $M(M-3)/2$
dimensions.


\end{document}